\documentclass[reprint,twocolumn,nopreprintnumbers,amsmath,amssymb,showpacs,aps,prl]{revtex4-2}

\usepackage{graphicx}
\usepackage{dcolumn}
\usepackage{bm}
\usepackage[dvipsnames]{xcolor}

\begin{document}

\preprint{APS/123-QED}

\title{\textcolor{black}{Separating water content from network dynamics in cell nuclei with Brillouin microscopy}}

\author{Lucie Vovard}
\author{Alexis Viel}
\author{Estelle Bastien}
\author{Lou-Anne Goutier}
\author{Ga\"etan Jardin\'e}
\author{Jérémie Margueritat}
\author{Sylvain Monnier}
\author{Thomas Dehoux}
 \email{thomas.dehoux@univ-lyon1.fr}
\affiliation{Universite Claude Bernard Lyon 1, CNRS, Institut Lumi\`ere Mati\`ere, UMR5306, F-69100 Villeurbanne, France}
\date{\today}

\begin{abstract}

\textcolor{black}{Probing forces, deformations–and generally speaking the mechanical properties of cells–is the hallmark of mechanobiology. In the last two decades many techniques have been developed to this end that are largely based on deforming the cells and measuring the reaction force. In cells, an alternative approach has been implemented mid 2010's, based on Brillouin Light Scattering (BLS) that produces a spectrum that can be interpreted as the response of the sample to an infinitesimal uniaxial compression at picosecond timescales. In all of these measurements, the response of the cell is quantified with a colloquial “stiffness” that encompasses both the contribution of load-bearing structures and volume changes, much to confusion. To clarify the interpretation of the hypersonic data obtained from BLS spectra, we vary the relative volume fraction of intracellular water and solid network by applying osmotic compressions to single cells. In the nucleus, we observe a non-linear increase in the sound velocity and attenuation with increasing osmotic pressure that we fit to a poroelastic model, providing an estimate of the friction coefficient between the water phase and the network. By comparing BLS data to volume measurements, our approach demonstrates clearly that BLS shift alone is mostly sensitive to water content while the additional analysis of the linewidth allows identifying the contribution of the biopolymer-based network dynamics in living cells.}

\end{abstract}

\maketitle

\section{\label{sec:level1}Introduction}

\textcolor{black}{The response of cells to forces and deformations plays a huge role in motility and structural integrity. Mechanical cues can also modulate polymer conformation within the cell and thus be translated into biochemical signals that support important molecular mechanisms. To understand these interactions, several tools to probe forces, deformations –and generally speaking the mechanical properties of cells –have being developed over the last two decades that are largely based on deforming the cells and measuring the reaction force. Among the many techniques, the most prevalent is probably AFM in which the surface of a cell is indented with a nanometer-scale tip. The analysis of the resulting force-displacement curves can provide a proxy for the shear modulus $G$ \cite{wu2018comparison}. Because water poorly supports shear, the general consensus is that non-zero values arise from the contribution of a meshwork of solid elements underneath the surface and within the cell, mostly the actomyosin complex in the cytoplasm and chromatin in the nucleus, that acts as a load-bearing structure \cite{haase2015investigating}.}

\textcolor{black}{In addition to mechanics, volume regulation is also key in maintaining important tissue functions, such as embryogenesis or wound healing. Perturbation of volume homeostasis, by external forces applied to the tissue or abnormal regulation, has also been associated with the development of degenerative diseases, senescence or cancer \cite{neurohr_excessive_2019, han_cell_2020}. Cell volume is controlled by osmotic regulation of the fluid content, via water efflux through the cell membrane. Modern implementations of AFM or micro-rheology have shown that, when compressing locally the cell with a tip, cell volume is reduced due to such water exchange, and that the mechanical modulus extracted from force-displacement curves actually correlates with cell volume  \cite{zhou_universal_2009, guo_cell_2017}. This suggests that, contrary to early interpretations of mechanical tests, changes in cell volume (or volume fraction of solid content) could be responsible for changes in the apparent cell stiffness.}

\textcolor{black}{There are thus two opposing visions that interpret cell deformability either in terms of load-bearing structures or in terms of volume fraction. However, these two approaches might be unified in dynamic measurements because the rate at which water escapes is controlled by the ability of the liquid to move through the meshwork of solid elements within the cell, and can be quantified by the so-called hydraulic conductivity in a classical Darcy experiment \cite{hu_using_2010}. In the cytoplasm, such an experiment has been reproduced by pushing on the cell with an AFM tip, and observing the time-dependent force relaxation. These observations were interpreted in the frame of the poroelastic theory (sketched in \textbf{Fig.\ref{fig:1}a}) established by Biot \cite{moeendarbary_cytoplasm_2013}. While a similar behavior has been posited for the nucleus, no definitive observation have been made \cite{wei_poroelasticity_2016}.}

\textcolor{black}{In cells, an alternative approach has been implemented mid 2010's, based on Brillouin Light Scattering (BLS) that uses the interaction of laser light with density fluctuations naturally present within the sample \cite{scarcelli_noncontact_2015}. The spectrum of the output light can be interpreted as the response of the sample to an infinitesimal uniaxial compression at picosecond timescales. The frequency shift and linewidth can be formally associated to the longitudinal sound velocity $v$ and attenuation $\gamma$, respectively. In cells, variations in frequency were initially interpreted as the viscoelastic response of the actin meshwork \cite{scarcelli_noncontact_2015}, trigerring a large interest from the mechanobiology community \cite{prevedel_brillouin_2019}. Considering that cells are highly hydrated materials, later studies pointed out the importance of water content \cite{wu_water_2018}, appealing for a mixing law description (sketched in \textbf{Fig.\ref{fig:1}a}). In multicellular tissues, BLS revealed a nonlinear increase in $v$ during osmotic regulation due to hydrodynamic coupling between the moving fluids and elastic network in tissues, pointing indirectly to poroelastic features that might stem at the single cell scale \cite{yan_probing_2022, margueritat_high-frequency_2019}. This confusion between water content, network elasticity and possible hydrodynamic coupling echoes that found in the broader mechanobiology community.}

\textcolor{black}{In all of these measurements, the response of the cell is quantified with a colloquial “stiffness” that encompasses both the contribution of load-bearing structures and changes in volume fraction, much to confusion. To clarify the interpretation of the hypersonic data obtained from BLS spectra,} we vary the relative volume fraction of intracellular water and solid network by applying osmotic compressions to single cells. In the nucleus, analysis of the spectra reveals a non-linear increase in the sound velocity and attenuation with increasing osmotic pressure. We demonstrate that, at low pressures, these features can be described by a simple mixture law owing to the high dilution of the nuclear solid contents. However, at larger pressures, we observe an excess acoustic attenuation due to the formation of a biopolymer network. To describe the contribution of a solid biopolymer-based network, we use a poroelastic model that gives a perfect match to the data across all pressures, providing an estimate of the friction coefficient between the water phase and the network. By comparing BLS data to volume measurements, our approach demonstrates clearly that Brillouin microscopy \textcolor{black}{allows separating water content from network viscoelasticity} in living cells. In the context of poroelasticity, this study provides a new framework to study water/network friction in cell nuclei and opens new avenues for studying molecular crowding in cells.

\section{\label{sec:level3}Results and discussions}
For illustrative purposes, we have chosen to study a colorectal cell line, HCT$116$, that is widely used in therapeutic research \cite{tammina_cytotoxicity_2017, jiang_long_2018, untereiner_drug_2018}. To vary the volumetric ratio between the non-permeant molecules and the intracellular fluids in a controlled manner, we use hyper-osmotic shocks. We rapidly increase the concentration of sucrose in the culture medium and, within seconds, the intracellular free water is drawn out of the cell through osmosis to equilibrate the solute concentrations on either sides of the cell membrane. In the cell, the nuclear envelope separating the cytoplasm from the nucleus is highly permeable to water, allowing free diffusion. This results in a well-established relationship between cell size and nuclear size under physiological conditions, known as the karyoplasmic or nucleocytoplasmic ratio \cite{cantwell_unravelling_2019}. This ratio is also conserved during osmotic compression since the osmotic pressure equilibrates in the two compartments \cite{finan_nonlinear_2009, pennacchio_n2fxm_2024}. As a result, upon compression, the volumes of the cell and the nucleus both decrease, therefore increasing the volume fraction of the solid content, $\phi$, corresponding mostly to ribosomes in the cytoplasm, and chromatin and proteins in the nucleus.

\begin{figure}[b]
\includegraphics{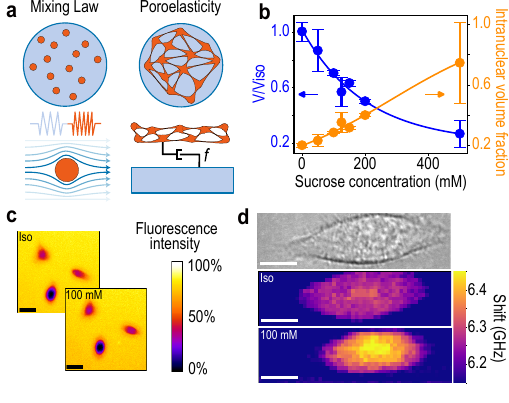}
\caption{\label{fig:1} \textbf{a}. Sketch of the mixing law (left) and poroelastic (right) models for a mixture of a solid network (orange) and a liquid phase (blue). In the mixing law model, the two components can be represented by springs in serie and the attenuation can be modeled by a Poiseuille flow. In the poroelastic model, the solid fraction forms a network permeated by the liquid phase. Friction, $f$, between the moving fluid and the elastic network couples their dynamics. \textbf{b}. Normalized cell volume measured with FXm (blue markers) and intranuclear solid volume fraction (orange markers) vs sucrose concentration (mM) and the corresponding fits (lines). \textbf{c}. FXm images of three cells at the isotonic condition and after a shock at $100$ mM. Scale bar: $20\;\mu$m. \textbf{d}. Bright-field image (top) and Brillouin frequency shift images at the isotonic condition (middle) and after a shock at $100$ mM (bottom). Scale bar: $10\;\mu$m.}
\end{figure}

To quantify this process, we measured cell volume with Fluorescence eXclusion microscopy (FXm) \cite{cadart_fluorescence_2017, zlotek-zlotkiewicz_optical_2015}. Briefly, the cells were suspended in a medium supplemented with a fluorescent dye attached to a large Dextran molecule that cannot cross the cell membrane, and they were inserted in a microfluidic chamber with a fixed height. Fluorescence is excluded from the cells, leading to a decrease in total fluorescence that is proportional to cell volume (see \textbf{Methods}). As an exemple, we show typical FXm images in \textbf{Fig.\ref{fig:1}c}.

Cell volume decreases with increasing sucrose concentration and reaches a limiting volume $V_{inact}$ due to osmotically inactive content formed by the solid fraction and bound water \cite{denysova_conclusions_2022}. To link cell volume $V$ to external sucrose concentration $C$, we fit these results with the following equation: $V/V_{iso}=(1-V_{inact}/V_{iso})\cdot e^{C/C_t}+V_{inact}/V_{iso}$. $V_{iso}$ is the isotonic cell volume, $C_t=200$ mM a characteristic concentration, and $V_{inact}/V_{iso}=0.2$ the relative osmotically inactive volume obtained from the best fit to the data (see \textbf{Fig.\ref{fig:1}b} blue line). This means that $~20\%$ of the cell volume is not compressible, as observed in previous reports \cite{xu_single-cell_2020, roffay_passive_2021}. Finally, due to osmotic equilibrium between the cytoplasm and the nucleus \cite{rollin_physical_2023}, we assume that the ratio between nuclear and cell volume is constant for a given cell line, as classically observed \cite{finan_nonlinear_2009, pennacchio_n2fxm_2024, wu_correlation_2022}. Therefore, using our results on cell volume compression, we can link the intranuclear volume fraction, $\phi$, and sucrose concentrations (see \textbf{Fig.\ref{fig:1}b} orange line) that we use in the remainder.

\begin{figure}[htpb]
\includegraphics{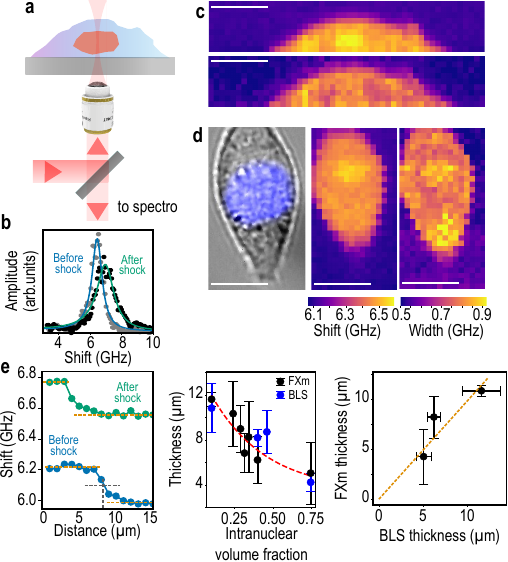}
\caption{\label{fig:3} \textbf{a.} BLS set-up. The laser light is focus inside the cell through a glass cover-slip. \textbf{b.} Typical BLS spectra in a cell (dots) and the corresponding fits to a Lorentzian function (lines), before (blue) and after (green) a shock at $500$mM. \textbf{c.} Brillouin frequency shift (top) and linewidth (bottom) images in the sagittal plane of a typical cell. Scale bar: $10\mu$m. \textbf{d.} Bright-field image of a cell with DNA labelled with Hoechst stain (left), and corresponding BLS images of the shift (center) and the linewidth (right). Scale bar: $10\mu$m. \textbf{e.} \textit{Left panel}: profiles across the nucleus of the BLS shift before (blue) and after (green) a shock at $500$mM. Orange dotted lines correspond to the cell and medium values. Grey line indicates the $\nu_B$-value where the cell thickness is measured. \textit{Centre panel}: Cell thickness measured with FXm (black) and that measured from BLS depth-profiles (blue) vs intranuclear solid volume fraction. \textit{Right panel}: Cell thickness measured with FXm vs. that obtained from BLS depth profiles (black dots). We plot the fit to a line with a slope of $0.98$ (orange dotted line).}
\end{figure}

Let us now measure the acoustic properties of the nucleus with BLS (see \textbf{Methods}). We use a backscattering geometry and a VIA-based spectrometer. We focus a $660$~nm laser beam through a 170 µm glass coverslip at the center of nucleus (see \textbf{Fig.\ref{fig:3}a}). For illustration, we plot typical spectra obtained before (blue) and after (green) shock in \textbf{Fig.\ref{fig:3}b} where we observe a clear increase in both the frequency shift and linewidth. In the remainder, we obtain the frequency shift, $\nu_B$, and linewidth, $\Gamma_B$, by fitting each spectrum to a Lorentzian function. The raw linewidth is corrected to account for instrumental broadening (see \textbf{Methods}). From $\nu_B$ and $\Gamma_B$ we obtain the acoustic velocity, $v$ and attenuation, respectively $\gamma$ \cite{pecora_dynamic_1985}:
\begin{align}
v &= 2\pi \nu_B/q\\
\gamma &=\pi \Gamma_B/v\label{eq-gammaB}
\end{align}
where $q$ is the optical wavenumber. \textcolor{black}{The attenuation is linked to the longitudinal viscosity using the Stokes-Einstein relation:}
\begin{equation}
    \textcolor{black}{\gamma = \frac{\omega^2}{2\rho v^3}} \label{ref-stokes-einstein}
\end{equation}
\textcolor{black}{where $\omega=2\pi\nu_B$ is the angular frequency}.

We scan the laser beam in the cell in a $2$D-raster-pattern to produce images of $\nu_B$ and $\Gamma_B$. We used a $\times60$ lens (NA $0.95$) and we measured a $2.1\;\mu$m axial (i.e. in-depth) point spread function (PSF), close to the theoretical diffraction limit. We illustrate the resolving power of the BLS microscope with images of $\nu_B$ and $\Gamma_B$ measured with a micron step in the transversal and sagittal planes of typical cells (\textbf{Fig.\ref{fig:3}c} and \textbf{\ref{fig:3}d}). For reference we plot the corresponding brightfield image with DNA labelled with Hoechst dye (blue). We also show in \textbf{Fig.\ref{fig:1}d} BLS images of the shift before and after shock (at the same sucrose concentration used in \textbf{Fig.\ref{fig:1}c}).

Yet from FXm data, we see that the thickness of the nucleus can be as small as $\sim4\mu$m at the highest sucrose concentrations. To clearly isolate the contribution of the nucleus to the BLS shift, we plot $\nu_B$ measured across the thickness of the cell vs distance from the glass/cell interface (see \textbf{Methods}) in \textbf{Fig.\ref{fig:3}e} (left) before (blue) and after (green) shock. Before shock, we clearly identify the cell nucleus with a shift at $\sim6.2$ GHz, while the surrounding medium shows a shift of $\sim6$ GHz (orange dotted lines). From such BLS-depth profiles we determine the nucleus thickness as the depth value where $\nu_B$ drops by $50\%$ (see gray dotted lines on \textbf{Fig.\ref{fig:3}e} (left)). We compare these values with the thickness obtained from FXm in \textbf{Fig.\ref{fig:3}e} (middle). We observe a significant reduction in the nuclear thickness, consistent with data from FXm. We also plot thickness obtained from FXm vs thickness obtained from BLS profiles in \textbf{Fig.\ref{fig:3}e} (right), and a fit to a line with a slope $0.98$, confirming the precision of the BLS measurements. \textcolor{black}{Because we didn’t observe any broadening in the Rayleigh peak, we assume multiple optical scattering plays no role in the broadening of the Brillouin peaks.}

\begin{figure}[t]
\includegraphics{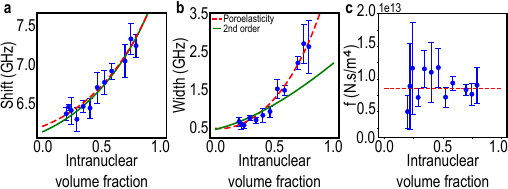}
\caption{\label{fig:4} \textbf{a} BLS shift vs intranuclear volume fraction and fit to the isostress (green line) and poroelastic (red dotted line) models. \textbf{b} Linewidth vs intranuclear volume fraction, with fit second-order correction (green) of the viscosity and to the poroelastic model (red dotted line) calculated from Eq. \ref{eq-gamma-poro} with $f=8.5\times10^{12}$ N.s.m$^4$ (see red dotted line in \textbf{Fig. \ref{fig:4}c}). \textbf{c.} Friction $f$ obtained from the measured linewidth plotted in \textbf{Fig. \ref{fig:4}b} and Eq. \ref{eq-gamma-poro}. The mean $f=8.5\times10^{12}$ N.s.m$^4$, used for the fit to poroelastic model in \textbf{Fig. \ref{fig:4}b}, is indicated by a red dotted line. We probed between $20$ and $40$ cells for each volume fraction.}
\end{figure}

From the BLS-profiles shown in \textbf{Fig.\ref{fig:3}e} we identify the maximum values for the frequency shift and the linewidth, $\nu_B$ and $\Gamma_B$, respectively, as the signature of the nucleus. We plot $\nu_B$ and $\Gamma_B$ vs intranuclear volume fraction in \textbf{Fig.\ref{fig:4}a} and \textbf{Fig.\ref{fig:4}b} (blue markers), and observe a significant increase in both quantities with increasing volume fraction. To explain this observation, by analogy with polymer solutions \cite{brown_dynamics_1991}, let us first describe the nucleus as a two-phase system composed of a fluid part (water) and a solid part made up by non-permeant components that cannot cross the membrane (i.e. mostly chromatin and other polymers, large sugars...), with a volume fraction $\phi$. With this description, a simple static mixing law allows determining the effective mass density:
\begin{equation}\label{rho_eff}
    \rho^*=\phi\rho_s+(1-\phi)\rho_f
\end{equation}
with $\rho_f$ and $\rho_s$ the mass densities of the fluid and solid parts, respectively. On the other hand, calculation of the effective longitudinal modulus requires to consider the dynamics of the sample. 

First, we rely on a description widely used in the Brillouin community, assuming that water content dominates the response in cells. We consider that the solid part of the nucleus is made up of independent inclusions without any hydrodynamic interactions between them, as is sketched in \textbf{Fig.\ref{fig:1}a}, similar to a dilute suspension. In this approach, one can use a mixing law, such as the ubiquitous isostress model: when compressed, the more compressible phase deforms the most (in this case the solid inclusions) and the stress is homogeneously distributed, as if the two phases were described by two springs in series. In this situation, the effective longitudinal modulus, $M^*$, is calculated using the Reuss average of the compliancies \cite{gadalla_transverse_2014},
\begin{equation}
    M^*\textcolor{black}{(\phi)}=\frac{1}{\frac{\phi}{M_s}+\frac{(1-\phi)}{M_f}} \label{Meff_isostrain}
\end{equation}
with $M_s$ and $M_f$ the longitudinal moduli of the solid and fluid parts, respectively. Using typical values $\rho_s=1350$ kg/m$^3$ \cite{gorisch_mobility_2005, howard_mechanics_2001} and $n^*=1.35$ \cite{schurmann_cell_2016}\footnote{In cells the ratio between the increase of refractive index, $\delta n$, and the increase in concentration, $\Delta c$, is $\Delta n/\Delta c=0.0019$, much smaller than the increase in the sound velocity, $\Delta V$, with typical values $\Delta V/\Delta c=7$ m/s in cells \cite{yan_probing_2022}. We can thus neglect it.}, we calculate \textcolor{black}{$\nu_B^*=2n^*v^*/\lambda$}, with $v^*=\sqrt{M^*/\rho^*}$ the effective sound velocity (Wood's law). We obtain $M_s=5.3$ GPa from the best fit to the data, a value close to that reported in gelatin gels \cite{bailey_viscoelastic_2020}. We obtain a very good match (see \textbf{Fig.\ref{fig:4}a}, green line), demonstrating the suitability of the isostress model to describe $\nu_B$ in the cell nucleus in this range of osmotic compressions.

To obtain the \textcolor{black}{effective linewidth using Eqs. \ref{eq-gammaB} and \ref{ref-stokes-einstein}, we need to calculate the effective longitudinal viscosity, $\eta_L^*$. To do so,} one can also turn to mixing laws to predict the effective shear viscosity of the suspension. A classical approach is based on the description of a flow around a single spherical inclusion (see \textbf{Fig.\ref{fig:1}a}) that predicts a linear dependence of the shear viscosity with $\phi$ \cite{brady_bulk_2006}. Second-order corrections can be implemented to better describe dilute monodisperse suspensions of rigid spheres:\footnote{\textcolor{black}{Equation \ref{Einstein} is obtained by combining the effective shear viscosity, $\eta_S^* = \eta_S^f\left(1+\frac{5}{2}\phi+\frac{31}{5}\phi^2\right)$, and the effective bulk viscosity, $\kappa*=\kappa^f+\frac{4}{3}\eta_S^f\phi$}, in $\eta_L^*=\kappa^*+\frac{4}{3}\eta_S^*\phi$.}
\begin{equation} \label{Einstein}
    \textcolor{black}{\eta_L^* = \kappa^f+\eta_S^f\left(\frac{4}{3}+\frac{14}{3}\phi+\frac{124}{15}\phi^2\right),}
\end{equation}
where $\eta_S^f=0.65$ mPa.s is the shear viscosity of the fluid (assumed to be water at $37^\circ$C) \cite{holmes_temperature_2011}, \textcolor{black}{and $\kappa^f$ is the bulk viscosity of the fluid part}. \footnote{\textcolor{black}{Note that using $\eta_S^f$ as a free parameter in the fitting procedure, the first order function produces aberrant trends while the second order function only fits parts of the range at the cost of unrealistic viscosity values.}} We use this simplistic description to calculate the linewidth, \textcolor{black}{$\Gamma_B^*$}, and compare it to the data in \textbf{Fig.\ref{fig:4}b} (green line). While this approach provides a reasonably good fit up to $\phi=0.4$, we see that it fails to explain the non-linear behavior we observe at higher volume fractions. Note that this model has been derived for frequencies lower than the fast relaxation of water. More intricate corrections can be implemented to account for inter-particle hydrodynamic interactions that can occur upon increased packing or increased frequencies \cite{illibauer2024diagnostic}, which, ultimately, emulates the existence of solid network.

\textcolor{black}{In the face of the evidence of the existence of network bathed in a liquid, by analogy with the application of BLS to gels since the 70's \cite{tanaka_spectrum_1973}, and because there is an emerging consensus that the nucleus behaves as a poroelastic material \cite{moeendarbary_cytoplasm_2013}, it is tempting to use a model based on Biot theory of acoustics in fluid-saturated porous solids \cite{hosea_elastodynamics_1986}. In highly hydrated materials, although a clear demonstration would require accessibility to additional scattering wavevectors as shown in previous studies in polymer networks \cite{filippidi2024}, it is reasonable to consider that the relaxation processes in water occur beyond the GHz regime. BLS hence probes the dynamic response of the network and its hydrodynamic interactions with the solvent, revealing poroelastic behavior without ambiguity.}

In such an approach, the sample is described as a solid network formed by the non-permeant material, invaded by the intranuclear fluid (see \textbf{Fig.\ref{fig:1}a}). The network has longitudinal and shear moduli, $M_n$ and $G_n$, respectively, distinct from those of the fluid or of the solid, driven by how the meshwork of individual solid inclusions is assembled. Note that $M_n$ is different from $M_s$. The acoustic displacement of the network, $u_n$, and the fluid, $u_f$ are modelled by linearized equations of motion, coupled by a damping term, $-f(\dot u_n - \dot u_f)$, describing the friction, $f$, resulting from the fluid moving relative to the network \cite{tanaka_spectrum_1973, bacri_etude_1979}. From this system of coupled equations, one can obtain the dispersion equation, and identify the velocity $v_p$ and attenuation $\gamma_p$ of the propagating modes from the real and imaginary parts of the roots, respectively. In the so-called gel limit that is usually considered in soft matter, in which the frame is supposed to be weak (with $M_n, G_n \ll M_f, M_s$), and considering relatively large friction values, such that $f\gg \rho^*\omega$, we obtain \cite{bacri_etude_1979, hosea_elastodynamics_1986}
\begin{eqnarray} 
    v_{p}&=&v^*\left[1+\frac{1}{2}\frac{M_{n}(1-M^*/M_s)^2}{M^*}\right] \label{poroV}\\
    \gamma_{p} &=& \frac{\omega^2}{2\rho^*v^*}\left[\frac{\textcolor{black}{\eta_L^f}}{v^{*2}}+\frac{(\rho_s-\rho_f)^2}{f}\phi^2\right] \label{eq-gamma-poro}
\end{eqnarray}
where $\rho^*$ and $M^*$ are described by Eqs. \ref{rho_eff} and \ref{Meff_isostrain}, respectively, and $\eta_L^f=\kappa^f+\frac{4}{3}\eta_S^f=5$ mPa.s (water at $37^\circ$C \cite{holmes_temperature_2011}) is the longitudinal viscosity. \textcolor{black}{Here we took $M_n=0.1M_f$ to satisfy the weak frame approximation.}

In the limit where $\phi\rightarrow0$ in Eq. \ref{poroV}, $v_p$ approaches $v^*$ plus a small positive correction due to the to non-zero stiffness of the frame. Eq. \ref{poroV} therefore only yields a small improvement to the already excellent fit to $\nu_B$-data provided by the simple mixture described earlier (see \textbf{Fig.\ref{fig:4}a}, red dotted line). It is therefore not decisive to consider poroelasticity for the description of $v$, as discussed in the past for gels\cite{hosea_elastodynamics_1986}. In Eq. \ref{eq-gamma-poro} however, we recognize a new damping term scaling in $\phi^2/f$. To examinate this contribution, we plot $f$ obtained from our data in \textbf{Fig.\ref{fig:4}c} using Eqs. \ref{eq-gammaB} and \ref{eq-gamma-poro}. We obtain a largely constant $f\approx8.5\times10^{-12}$ N.s/m$^4$, a value comparable to those found in soft polyacrylamide gels \cite{tanaka_spectrum_1973}. Using this constant $f$-value in Eq.\ref{eq-gamma-poro}, we calculate the attenuation $\gamma_p$. From $\gamma_p$ we obtain \textcolor{black}{$\Gamma^p_B$} using Eq.\ref{eq-gammaB} and plot it in \textbf{Fig.\ref{fig:4}b} (red dotted line). We see a very good agreement. This comparison suggests the relevance of the poroelastic theory in the weak frame limit to describe osmotic compressions of cell nuclei, here used with only $f$ as an adjustable parameter. 

While such an agreement is in line with early studies on soft gels \cite{tanaka_spectrum_1973, bacri_etude_1979} that are classically considered as proxies for cellular material \cite{wu_water_2018, rodriguez-lopez_network_2024}, it does not align with later interpretations of the Biot theory \cite{hosea_elastodynamics_1986}. In particular, the friction, $f=\eta_L^f/k$, can be written as the ratio between the fluid viscosity and the permeability of the flow channels, $k$ \cite{johnson_elastodynamics_1982}. Thus high $f$-values are interpreted as originating from either a large viscosity or low permeability that locks the network and fluid together, due to the viscous skin depth becoming larger than the effective pore size. In an attempt to estimate the effective pore size, one can assume that the flow channels are cylindrical tubes of radius $r$ \cite{biot_theory_1956}, leading to a $\phi$-dependent expression for permeability, $k=(1-\phi)r^2/8$, translating the fact that the network becomes more dense as it is compressed. In contrast, our finding of a constant friction coefficient, \textcolor{black}{as previously observed in gels \cite{johnson_elastodynamics_1982},} suggests that the damping mechanisms involve more intricate channel geometries, with possibly a non-negligible tortuosity.

\textcolor{black}{Alternatively, it might be relevant to consider the contribution of other coupling phenomena such as inertial drag of the fluid upon network motion \cite{hughes_estimation_2003}, or separating the frictional influence of the free and polymer-bound water \cite{chiarelli_acoustic_2009}. To go further, it would be important to validate the poroelastic nature of the nucleus with an alternative technique that would quantify the flow of liquid due to a hydrostatic pressure, as in a classical Darcy experiment, but at the single cell scale. Although the poroelastic model give a good description of the linewidth during osmotic shocks, other models, e. g. based on multiple scattering, could be considered. In scenarii where water content is modulated by processes other than osmotic regulations, alternative descriptions of the hydrodynamics of the polymers might account for network formation through increased packing or increased frequencies, or the formation of crosslinks in specific physiological conditions.}

\textcolor{black}{In conclusion, our work shows that in cell nuclei, concerning the frequency shift, mixing laws (i.e. isostress model) and poroelastic modelling produce indiscernibly good fits to the data, even at high molecular concentrations, in line with previous conclusions that water content dominates the spectral response. Looking at the linewidth however, mixing laws only provide a good description in the dilute regime, as they are designed to. Above a certain concentration threshold, it is necessary to account for the hydrodynamic interactions between the solid molecules that form a soft network. This observation demonstrates that BLS measurements can distinguish changes in volume fraction (probed by the shift) from stiffening of the polymer network (probed by the linewidth) due to the reduction of water content. This observation calls for caution when attempting to extract mechanobiology readout (e.g. stiffness) from BLS spectral features, especially from the frequency shift alone: variations in the shift or linewidth cannot be univocally ascribed to modifications in load-bearing structures without precise knowledge of the cell volume fraction. Looking back at similar observations of a correlation between the shear moduli --probed by AFM or micro-rheology --and cell volume \cite{zhou_universal_2009, guo_cell_2017}, the same precaution presumably applies to any approach that measures an elastic modulus.}

\section{\label{sec:level4}Material and Methods}
\subsubsection{Cell culture}
The HCT$116$ Wild Type (WT) cell line was purchased from the Amercian Type Culture Collection (ATCC). HCT$116$ WT cells are maintained in Dulbeco’s Modified Eadle Medium (DMEM) supplemented with $10\%$ fetal bovine serum (FBS; PANBiotech), $1\%$ penicillin/streptomycin (P/S) and glutamax (Life Technologies), at $37^\circ$C under a $5\%$ CO2 and at $95\%$ relative humidity. For fluorescent imaging, after a PBS wash, cells were incubated with Hoechst $33342$ (ThermoFischer) at a concentration of $10\;\mu$g/mL for $10$ minutes at $37^\circ$C. After incubation, cells were washed three times with PBS. Nuclear staining was visualized using a fluorescence microscope with a DAPI filter.

\subsubsection{Osmotic shock}
We induce osmotic pressure perturbations with solutions of sucrose at different concentrations (from $100$mM to $550$mM). The solutions were prepared by diluting sucrose powder (Sigma Aldrich) into imaging medium (DMEM supplemented with glutamax, $10\%$ FBS, $1\%$ P/S and $20$mM HEPES from Life Technologies). Culture medium was replaced by sucrose solutions at least $10$min before measurement to assure homogeneity in the petri dish. After every measurement, the osmolarity of the shock solutions was checked with a cryoscopic osmometer (Löser automatic typM $10/25\;\mu$L).

\subsubsection{Brillouin spectroscopy}  
We use a $660$m single mode laser coupled to an inverted life-science microscope (Nikon Eclipse Ti$2$-U) to focus and collect backscattering light with a objective lens $\times60$ (N.A. $0.95$). The microscope is enclosed in an environmental chamber to maintain the samples $37^\circ$C. For each condition, we probed between $20$ and $40$ cells with a LightMachinery VIPA spectrometer (HF-$8999$-PK-$660$-$15$GHz Rev.A). We used an acquisition time of $1$s for all BLS measurements with a laser power of $30$mW on the sample. Due to the spectrometer and aperture of the objective lens, we observe a spectral broadening of $180$ MHz that we measured on reference saline solutions of increasing concentration\cite{yan_evaluation_2020}. We subtracted this value from the linewidth data. We also measured the accuracy of the system and we found a precision of $30$ MHz for the shift and $25$ MHz for the linewidth. The point spread function (PSF) of the microscope was measured at $2.1\;\mu$m by monitoring the amplitude of the Brillouin of ethanol when crossing a glass/ethanol interface \cite{bacete_theseus1_2022}. 


\subsubsection{Data analysis}
Typical acquisition time is $1$ s per spectrum with $20$ mW at the sample. Each data point represent between $20$ and $40$ different cells, and the shocks were repeated $3$ times at each concentration ($N=3$). For the $z$-profiles, we acquire a spectrum every micron along the $z$-axis (perpendicular to coverslip surface) from below the glass/cell interface all way up above the cell in the culture medium. To determine the position of the glass/cell interface, we monitor the amplitude of the Rayleigh peak that is maximum at the interface. In highly compressed cells at high osmotic pressure, we sometimes observe $2$ Brillouin peaks corresponding to the cell and to the culture medium. In this situation, we fit the sum of two Lorentizian function (one of these is that of the medium that we have characterized at every sucrose concentration) to the spectrum.

\subsubsection{Fluorescence exclusion microscopy}

Briefly, cells were incubated in polydimethylsiloxane (PDMS) with medium supplemented with a fluorescent dye coupled to Dextran molecules ($10$ kDa) to prevent entry into cells.\cite{zlotek-zlotkiewicz_optical_2015}. Fluorescence is thus excluded by the cells, and volume is obtained by integrating the fluorescence intensity over the cell. Chips for volume measurements were made by pouring PDMS elastomer and curing agent ($1:10$, Sylgard $184$ from Dow Corning) on a mold and were cured for at least $2$ h at $65^\circ$C. Molds were made on a silicon wafer with SU-$8$ photoresist using classical photolithography techniques. $3$ mm inlets and outlets in the PDMS were punched, and the PDMS chips were cut to fit on glass coverslips. The coverslips were bonded onto $35$ petri dishes with Norland Optical Adhesive $81$ (Norland) for $1$ minute at UV $324$ nm. Chips were bonded to glass coverslips by exposure to air plasma for $30$ s and incubated with Poly-L-Lysine solution for $30$ min (Sigma-Aldritch) or no treatment was made. Chips can be used directly or stored in PBS at $+4^\circ$C for weeks. \\

One day before the experiments, PBS was replaced by medium and cells seeded in the chip. the day of the experiment, medium was changed to phenol-free medium supplemented with HEPES and complemented with fluorescent dextran. Imaging started within $10$ min after cell injection in order to prevent adhesion and thus, cells response to the shear stress generated by medium exchange. Acquisition was performed at $37^\circ$C in CO$_2$ independent medium (Life Technologies) supplemented with $1$ g/L Alexa$647$ Dextran ($10$ kDa; Thermo Fischer Scientific) on an epifluorescence microscope (Leica DMi$8$) with a $10\times$ objective (numerical aperture (NA) $0.3$).

\nocite{*}

\bibliography{apssamp}

\end{document}